# Effect of Strain and Diameter on Electronic and Charge Transport Properties of Indium Arsenide Nanowires

Pedram Razavi[1] and James C. Greer[2]

Tyndall National Institute, University College Cork, Lee Maltings, Dyke Parade, Cork, T12 R5CP, Ireland

University of Nottingham Ningbo New Materials Institute and Department of Electrical and Electronic Engineering, University of Nottingham Ningbo China, 199 Taikang East Road, Ningbo, 315100, China

*Abstract* — The impact of uni-axial compressive and tensile strain and diameter on the electronic band structure of indium arsenide (InAs) nanowires (NWs) is investigated using first principles calculations. Effective masses and band gaps are extracted from the electronic structure for relaxed and strained nanowires. Material properties are extracted and applied to determine charge transport through the NWs described within the effective mass approximation and by applying the non-equilibrium Green's function method. The transport calculations self-consistently solve the Schrödinger equation with open boundary conditions and Poisson's equation for the electrostatics. The device structure corresponds to a metal oxide semiconductor field effect transistor (MOSFET) with an InAs NW channel in a gate-all-around geometry. The channel cross sections are for highly scaled devices within a range of $3 \times 3$ nm$^2$ to $1 \times 1$ nm$^2$. Strain effects on the band structures and electrical performance are evaluated for different NW orientations and diameters by quantifying subthreshold swing and ON/OFF current ratio. Our results reveal for InAs NW transistors with critical dimensions of a few nanometer, the crystallographic orientation and quantum confinement effects dominate device behavior, nonetheless strain effects must be included to provide accurate predictions of transistor performance.

## 1. Introduction

Field-effect transistors (FETs) are anticipated to be manufactured with sub-7 nm critical dimensions within the next few years[1]. Electronic properties of materials at these length scales vary significantly with respect to their bulk values due to the dramatic increase in the surface-to-volume ratio and quantization effects arising at small critical dimensions[2,3]. Due to large quantum confinement effects in NWs, there is substantial band gap widening relative to the bulk energy gap. In addition to band gap widening, the direct or indirect nature of a semiconducting material can be altered due to band folding[4-6] leading to fundamentally different electrical and optical properties relative to the bulk. Improvement of the electrical performance and obtaining satisfactory electrical drive current in the highly-scaled MOSFETs requires technology boosters such as thin-body channels and strained nanowires, and as well may require high transport channel materials such as germanium and/or III-V compound materials[1,7]. A feature of III-V semiconductors is the possibility to control device dimensions, doping concentrations, and material composition while achieving band gap engineering during fabrication[8,9]. However when scaling any MOSFET architecture, a set of deleterious performance issues such as degradation of subthreshold swing (SS) and ON/OFF current ratio ($I_{ON}/I_{OFF}$) emerge which are collectively referred to as short-channel effects (SCEs)[10]. Nanowire structures with a gate electrode wrapped around the NW axis, known as the gate-all-around (GAA) configuration, enhances electrostatic control of the channel carriers and considerably mitigates against short-channel effects for highly scaled dimensions[1]. III-V materials maintain very high electron mobility together with good drive currents in NW field-effect-transistors (FETs)[11] and substantial progress on the fabrication of III-V NW devices on a Si substrate has been achieved in recent years[12]. However in devices scaled below 7 nm critical dimensions, extrapolations based on bulk and classical device concepts need to be tested[1]. The impact of strain in NWs can considerably affect device characteristics beyond what is observed at larger dimensions for physical properties such as the band gap and effective masses, and these effects are strongly dependent on the crystallographic orientation of the nanowire primary axis[13]. Moreover, strain engineering is well known to be capable of improving charge carrier mobility in semiconducting materials[13]. However, unlike the group IV semiconductors silicon and germanium as well as their alloys, there is less known about the effect of strain on the physical properties of III-V nanowires[4,6,14-17], particularly in the context of nanoelectronic devices. Recently, nanowires with diameters as small as 1 nm have been fabricated relying on recent advances in bottom up as well as top down fabrication techniques. Ultra scaled SiNWs with approximately 1 nm diameters with their oxide sheaths removed and replaced with hydrogen termination have been prepared and the band gap widening due to quantum confinement was reported [18]. More recently, InAs NWs with approximately a 2 nm diameter have been fabricated using a metalorganic chemical vapor deposition (MOCVD) via the vapor-liquid-solid (VLS) growth method with a gold nanoparticle catalysis[19]. To this end, the band structures of InAs nanowires are investigated with the goal of extracting accurate transport models for nanowire transistors[20] with cross sections in the range of approximately 3 ×3 nm$^2$ to 1×1 nm$^2$. These cross sectional areas are chosen for extracting the quantum confinement and strain effects relevant for modern nanowire designs for nanoelectronics. These ultra scaled critical dimensions are exactly the scale at which nanoelectronics manufacturers are currently exploring for the beyond 7 nm technology nodes[21-23] and it is thus timely to explore the electronic properties for nanowires with critical dimensions approaching atomic scale limits. Herein different wire orientations are considered and the effect of strain is also evaluated through the use of density functional theory (DFT) calculations. Quantum transport simulations are also performed to obtain the transfer characteristics of transistors with InAs nanowire channels and subthreshold swings (SS), and drive currents ($I_{ON}$) are extracted.

## 2. Methods

The DFT calculations are performed using a linear combination of atomic orbitals (LCAO) consisting of a polarized double-zeta basis set. A meta-generalized gradient approximation (meta-GGA) for the exchange-correlation functional in the DFT calculations is used[24-26]. The energy band gap of bulk InAs is relatively narrow at room temperature ($E_g$=0.354 eV)[20] and the well-known band gap underestimation of standard approximations to the exchange-correlation functionals such as the local density approximation (LDA) or generalized gradient approximation (GGA) make their use unsuitable for narrow gap semiconductors. The importance of an accurate treatment of the kinetic energy density for the calculation of band gaps in solids using density functional theory is also known[27]. Meta-GGA[28] functionals relate the exchange-correlation energy at each point not only to the local electronic density, but also to the electronic density gradient and the Kohn-Sham kinetic energy density. As a result, meta-GGA functionals can provide a more accurate band gap prediction, however at the expense of an empirical calibration. By fitting the "*c*"-parameter of the Tran and Blaha exchange–correlation functional[24], the energy band gap for bulk InAs is calibrated to the experimental value of 0.354eV [20]. By doing so we obtain an effective mass of 0.023 in good agreement with effective mass of 0.026 from ref. [19]. For all nanowire calculations, a Monkhorst-Pack sampling[29] with a

23×1×1 *k*-point grid and energy cut-off of 100 Hartree is used to generate the real-space grid and for bulk calculations a 13×13×13 *k*-point grid were used. The total electronic energy for each NW is minimized with respect to atomic positions. The atomic configurations are relaxed to a force of less than 0.01 eVÅ$^{-1}$atom$^{-1}$ and the resulting geometries are taken as the zero strain reference configurations.

Electron effective masses at conduction band minima are calculated using a parabolic approximation with a second derivative $\partial^2 E/\partial k^2$ determined using a 5 point stencil method[30]. Approximately square zinc-blend (ZB) InAs nanowires are generated with wire orientations of [100], [110] and [111] with cross-sectional areas of approximately 3×3 nm$^2$ to 1×1 nm$^2$. It is important to note that cross sectional shape can also influence electronic properties of the NWs, although not considered in the present calculations and is left for future studies. The NW surfaces are passivated using pseudo-hydrogen atoms[31] to ensure saturation of all surface dangling bonds thus avoiding issues relating to specific surface chemistry and bonding on the electronic properties of the NWs – in this sense the calculations provide a reference for the ideal surface termination. Fig. 1 shows typical cross sectional views of the relaxed InAs NWs for different wire orientations used in our study.

## 3. Electronic structure results

To determine the effect of the wire cross section and orientation on characteristics of InAs nanowires at low critical dimensions, the band gaps and effective masses are extracted from the meta-GGA electronic structures and plotted in fig. 2a; these are the two critical parameters for determining transistor performance as the band gap widening due to quantum confinement governs thermal and tunneling properties thus influencing ON/OFF current ratios, and electron effective masses can be considered as a first approximation to the changes in electron mobility, with smaller effective masses correlating to higher electron mobility. To examine the correction provided by the meta-GGA functional, the values extracted from the LDA band structures are also plotted in fig. 2a for comparison.

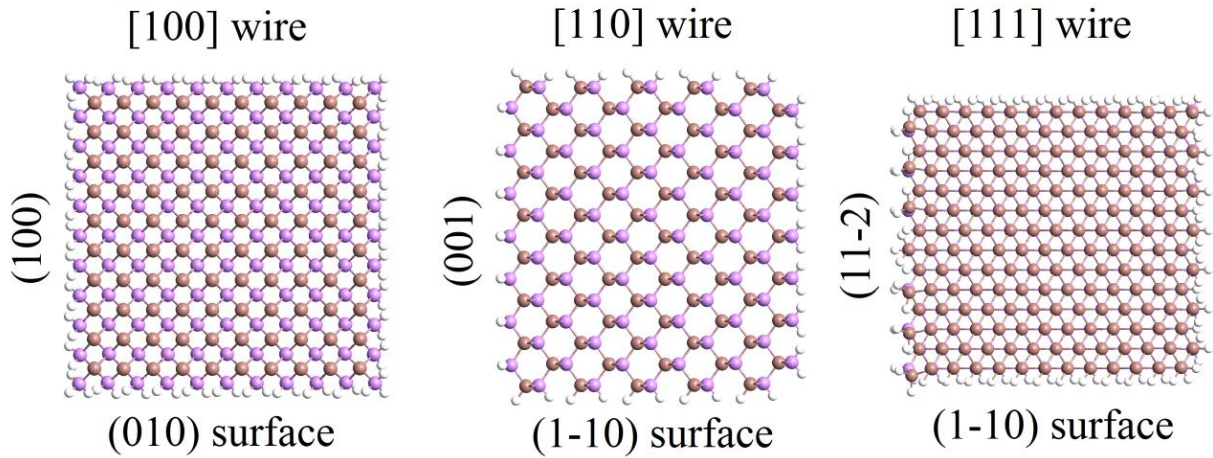

*Fig. 1.* Cross-section of InAs NWs for [100], [110], and [111] wire orientations composed of As (pink), In (brown) and pseudo-H (white) atoms. Note that pseudo-hydrogens of appropriate charge are chosen for each surface atom to provide a defect free surface.

It can be seen that unlike bulk InAs which has isotropic effective mass at the Γ-point, for small diameter InAs nanowires the effective masses at Γ are not isotropic and wires with different orientations have different effective masses due to the large quantum confinement effects normal to the NW principal axis. Fig. 2b shows the band structures of 1 nm and 3 nm nanowires with different wire orientation in which the effect of quantum confinement on the band gap and the effective masses can be seen. The band gap and effective masses are highly dependent on nanowire diameter and orientation which are the first indication of the impact these physical parameters for the electrical and optical properties of nanoscale NWs.

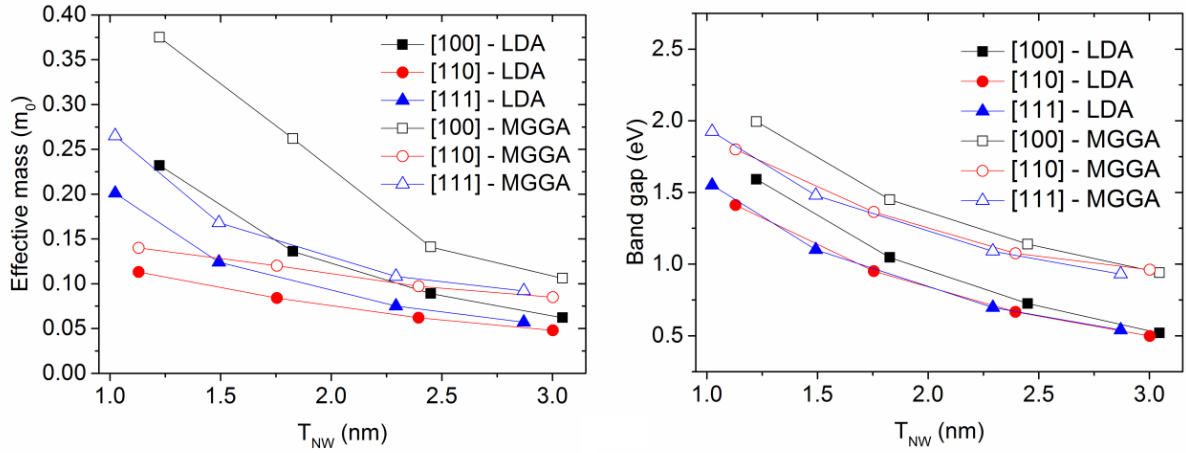

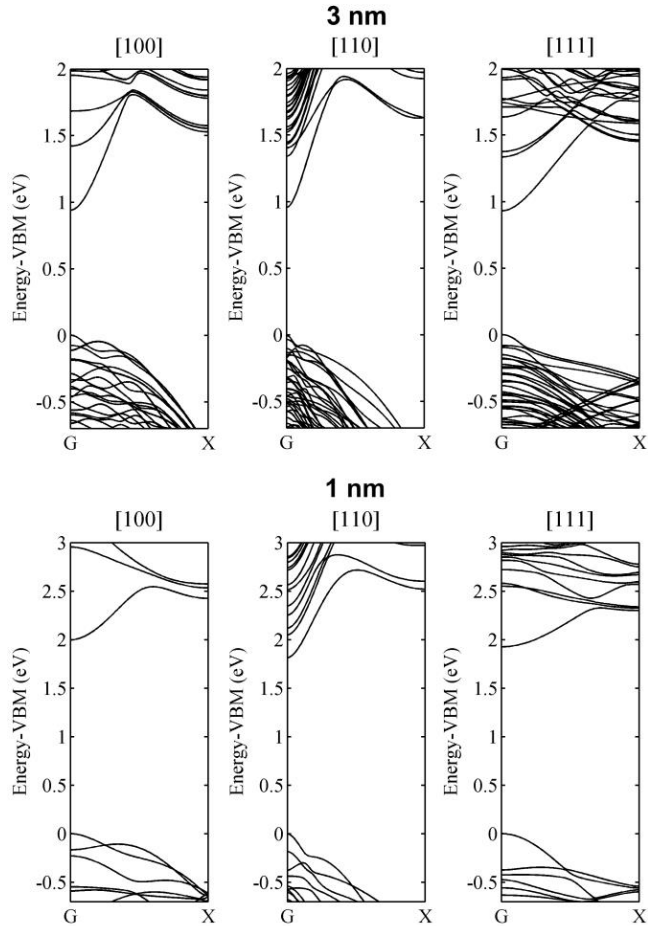

*Fig. 2.* Comparison of (a) band gaps and effective masses and (b) band structure of approximately 1×1 nm$^2$ and 3×3 nm$^2$ InAs NWs with different cross-section diameters and wire orientations. Note that the Meta-GGA approximation as anticipated predicts a larger band gap than the LDA calculations, as well, the effective masses tend to be slightly

larger within the Meta-GGA calculations. $T_{NW}$ dentoes the length of a side of the approximately square InAs NWs. G and X indicates the Γ point and the zone edge $\pi/a$. Note that unlike what it may look like from the curvature of these band structures, due to the larger lattice constant of the (111)-oriented wire the calculated effective mass will be smaller than the (100)-oriented wire as $\pi/a$ would be smaller. (Lattice constant of relaxed nanowires with no strain are given in Table. 1)

Table. 1. Lattice constants (Å) of relaxed 1 nm and 3 nm InAs nanowires with no strain

| 1 nm [100] | 3 nm [100] | 1 nm [110] | 3 nm [110] | 1 nm [111] | 3 nm [111] |
|---|---|---|---|---|---|
| 6.083 | 6.10 | 4.303 | 4.319 | 10.517 | 10.464 |

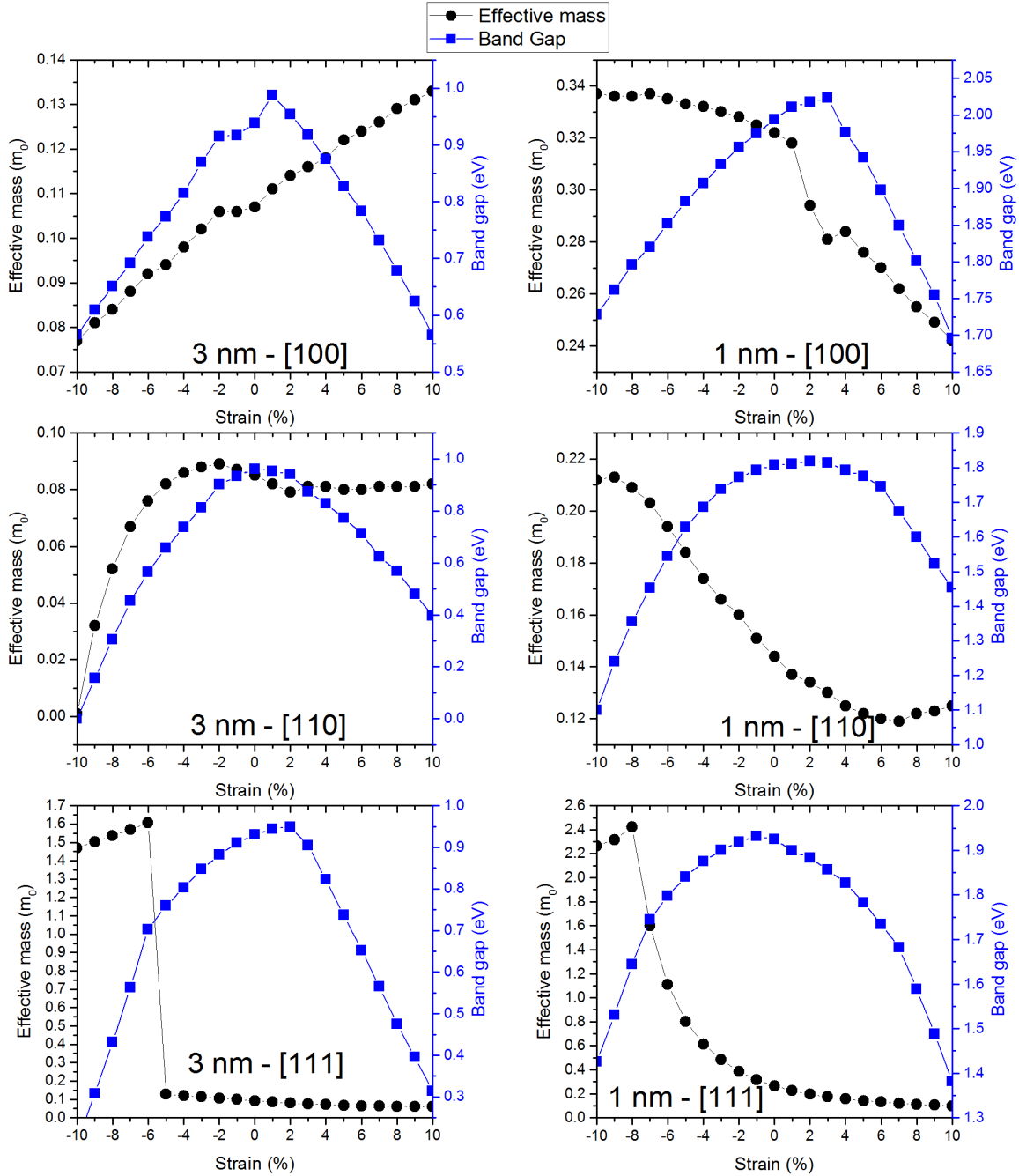

*Fig. 3.* Comparison of band gaps and effective masses for InAs NWs with strain up to ±10%.

The effect of strain on the band structure of InAs nanowires is investigated by applying tensile and compressive uni-axial stress along the NW principal axe. After the simulation cell parameter in the direction of the NW axis, the atomic positions are relaxed and lattice constants are varied to obtain strains of up to ±10%. Fig. 3 illustrates the effect of applying strain to the NWs on the band gap energies and effective masses. Clearly the impact of tensile and compressive strains is significant and should not be neglected in calculation of effective mass electronic structure properties. It is noted that the [111]-orientated NWs are found to be more prone to strain than [100]- and [110]- oriented NWs which can be seen in fig. 4.

Fig. 4 illustrates that compressive strain lowers the heavy hole subbands for all NWs orientations and lowers the heavier off-center valleys of the conduction band. Heavier off-center valleys may become the ground state valleys resulting in transition of direct into the indirect bandgap NWs. This transition can be seen for 3 nm [111]-oriented NWs with strain above 6% and for 1 nm [111]-oriented NWs with strains above 8%. On the contrary, tensile strain raises heavy hole subbands while lowering the light valleys at Γ with respect to the heavier off-center valleys. If the heavy hole subbands cross the light hole subbands, a transition from direct to indirect bandgap semiconductor NWs occurs. This mechanism is seen for the cases of the 3 nm and 1 nm [110]-oriented NWs with strains above 2 % and 4%, respectively.

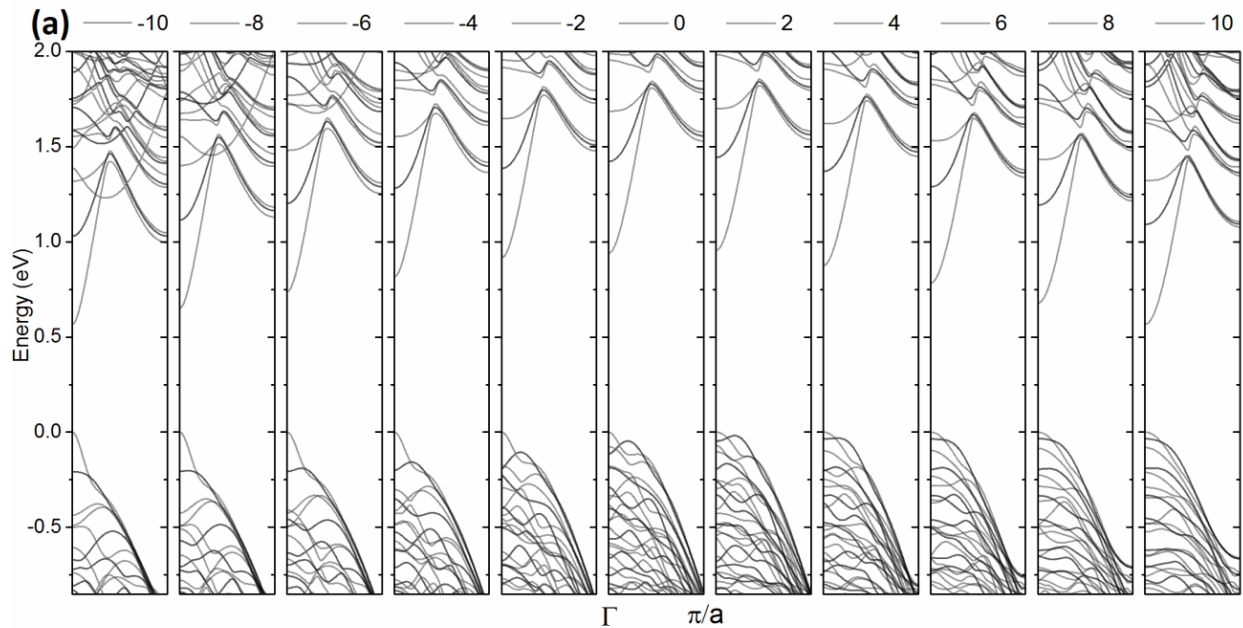

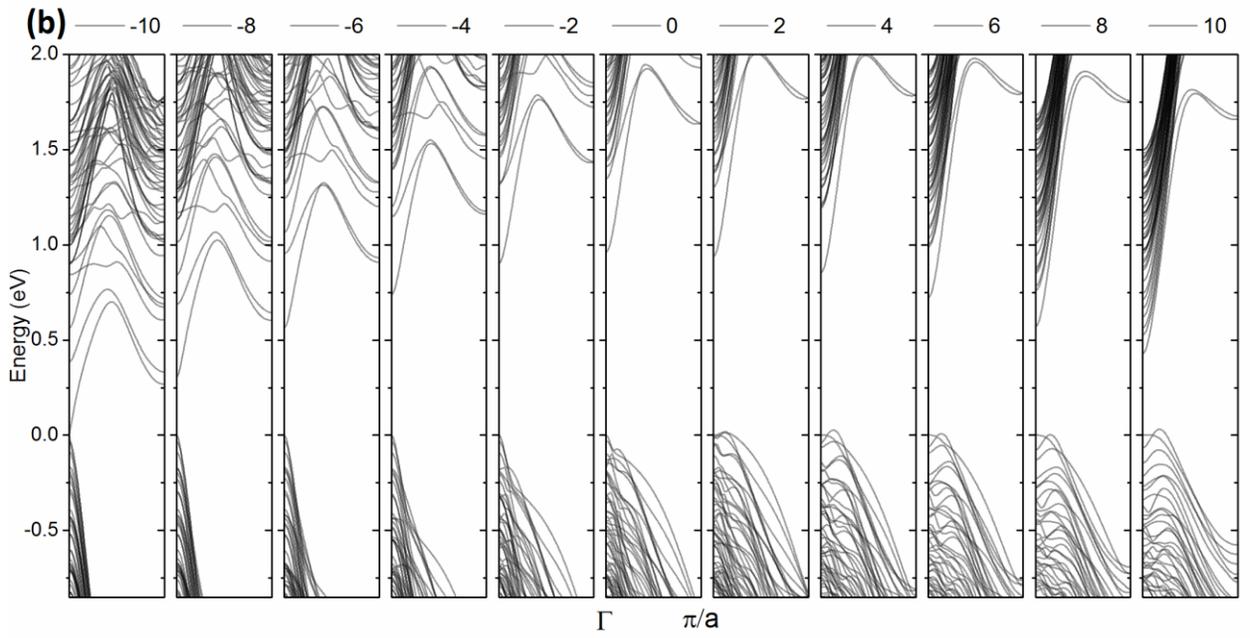

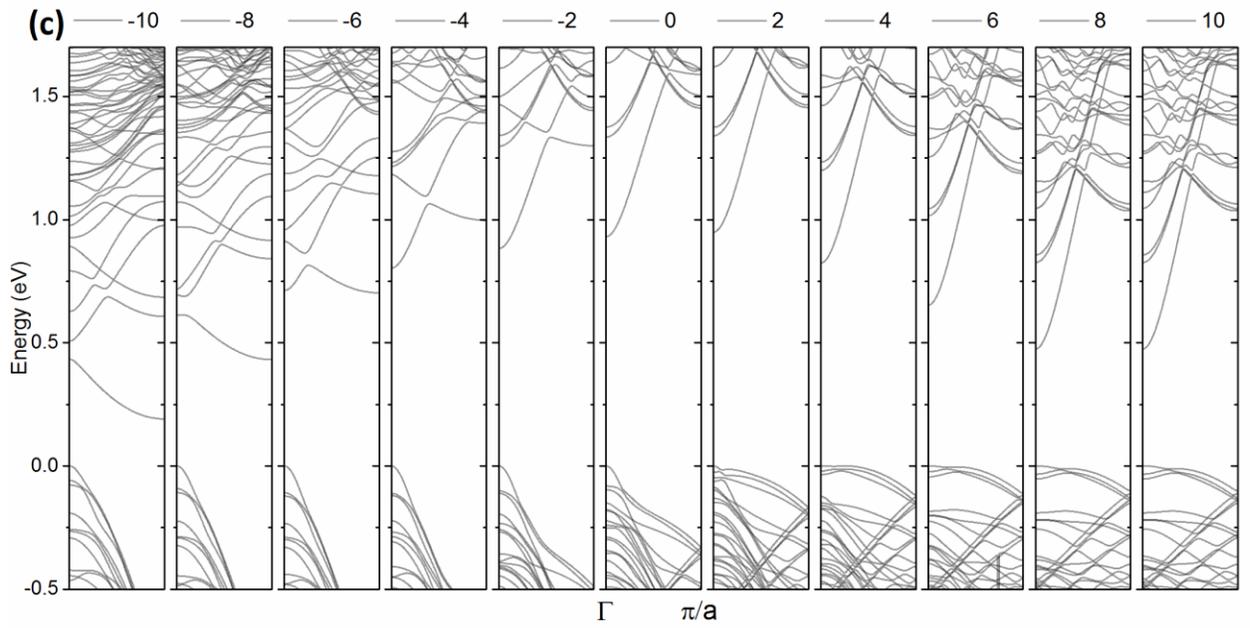

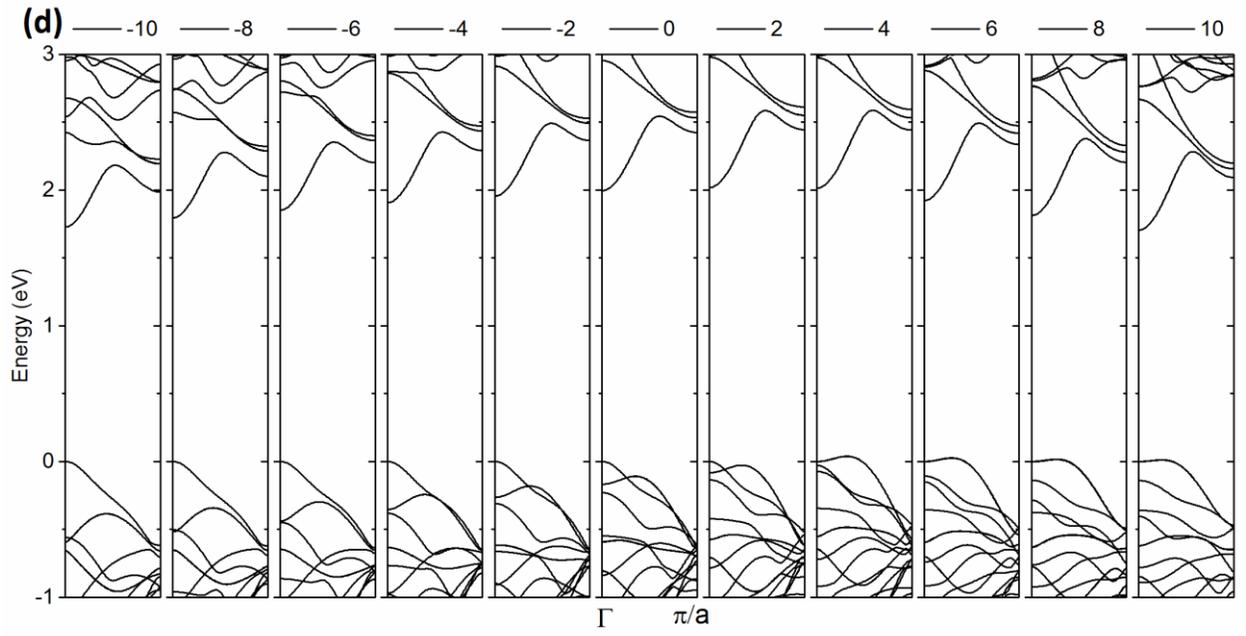

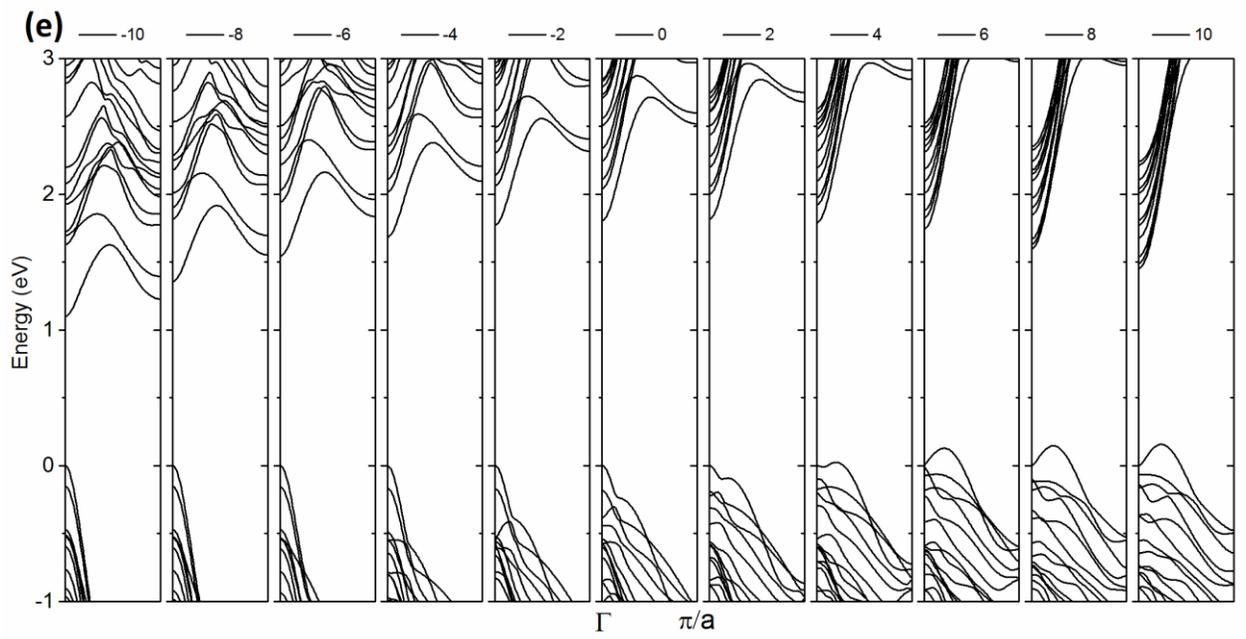

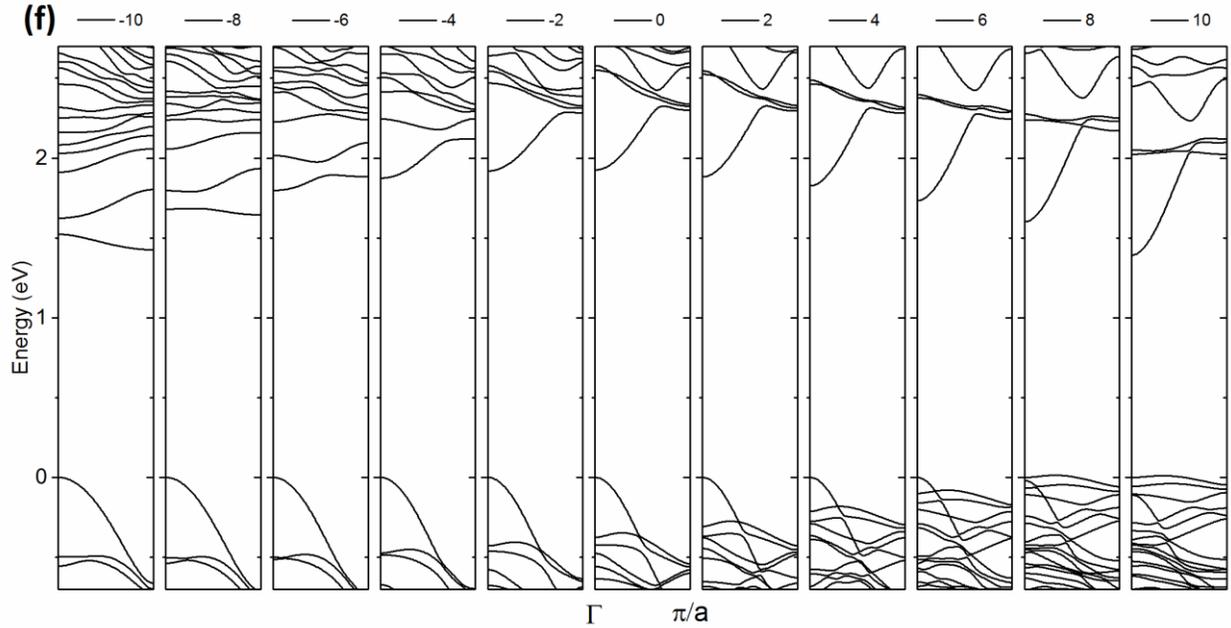

*Fig. 4* Band structures of strained (a) 3 nm [100]-, (b) 3 nm [110]-, (c) 3 nm [111]-, (d) 1 nm [100]-, (e) 1 nm [110]-, (f) 1 nm [111]-oriented InAs NWs (up to ±10%). The numbers of each figure give the applied strain $\Delta l/l \times 100\%$.

The combined results reveal how wire orientation, diameter and strain affect electrical properties of InAs nanowires with critical dimensions on the order of a few nanometer. To investigat the impact of these electronic structure changes affect InAs NW transistor characteristics, charge transport results are presented in the following section.

## 4. Charge transport results

The I-V characteristics of InAs gate-all-around nanowire transistors with tensile and compressive uni-axial strains up to ±6%, respectively, are calculated based on a self-consistent solution of the Schrodinger and Poisson equations via the non-equilibrium Green's function approach in the effective mass approximation and within the ballistic transport regime. Details of the method and implementation of the quantum transport calculations are given elsewhere[1]. The schematic of the nanowire cross-section in the longitudinal direction is depicted in fig. 5. The NW effective masses and band energies of the first eight conduction subbands are extracted from the DFT-determined electronic structure calculations using the meta-GGA approximation presented in the previous section with the electronic structure results defining the effective mass Hamiltonian. The subthreshold swing (SS) measures the rate of current increase with gate voltage below threshold, which is of particular importance to assess low power performance in ultra-scaled devices[32]. The SS along with the ON/OFF current ratio ($I_{on}/I_{off}$)$_{OFF}$ are calculated as the relative performance indicators for the different InAs transistors evaluated in the charge transport calculations. The supply voltage is fixed at $V_{dd}$=0.65 V and the gate work function is chosen for each nanowire transistor such that the off-current is 10 pA/µm at $V_{gs}$= 0 V, which is suitable target for low standby power technologies[1]. The length of the gate and source/drain regions are 10 and 15 nm, respectively, and the source/drain junctions are assumed to be abrupt. A gate-all-around configuration is assumed. The gate oxide thicknesses are 1 nm and the dielectric constants is chosen as 3.9 corresponding to silicon dioxide. Explicit gate tunneling currents are not included in the simulations, however the gate coupling can be maintained by scaling to a thicker oxide layer with a corresponding increase in the gate dielectric constant to suppress tunneling for a physical device. Uniform doping concentrations in the source/drain regions of $10^{20}$ cm$^{-3}$ for approximately 3×3 nm$^2$ wires and 6x$10^{20}$ cm$^{-3}$ for the approximate 1.2×1.2 nm$^2$ cross sections are chosen to obtain a similar number of dopant atoms in source/drain regions for both structures. This results in scaling of the dopant concentration with nanowire cross section.

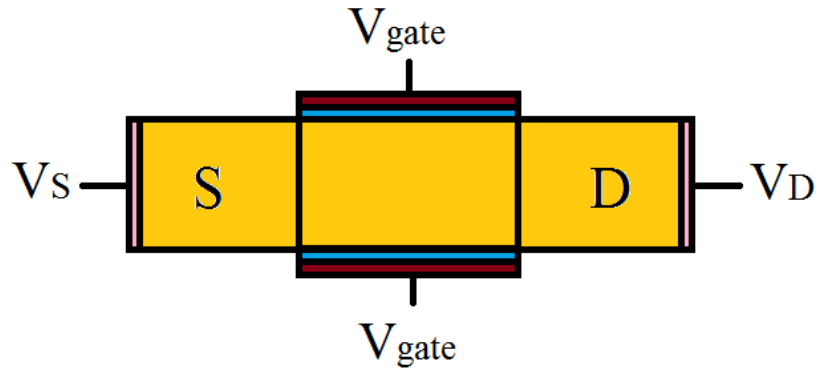

*Fig. 5* Schematic of the nanowire cross-section in the longitudinal direction.

Fig. 6 shows the subthreshold swing and $I_{on}/I_{off}$ for different InAs nanowire transistors. From fig. 6, it is found that at a fixed gate length of 10 nm, as the cross-section decreases the subthreshold swing approaches the theoretical limit for room temperature operation of 59.6 mV/decade. This is a direct consequence of the better electrostatic control of the gate over charge carriers in devices with smaller cross-sections as can be explained by the natural length for multi-gate architectures[10] which is defined by the following expression

$$\lambda = \sqrt{\frac{\varepsilon_s}{4\varepsilon_{ox}} t_s t_{ox}} \qquad (1)$$

where $\varepsilon_{ox}$, $\varepsilon_s$, $t_{ox}$ and $t_s$ are the permittivity of the gate oxide, permittivity of the wire material, gate oxide thickness and nanowire thickness, respectively. The natural length is a parameter which represents the extension of the electric field into the channel region.[1] the ratio of effective gate length to the natural length should be large enough for transistors to minimize SCEs. From the expression it is clear that decreasing the nanowire thickness can minimize the SCEs thereby improving the SS. In addition, the results can be explained by the findings for the effective masses in the transport direction (fig. 2a) with a larger effective mass in the transport direction reducing source-to-drain tunneling and thereby improving the SS. It is also found that 3 nm [111]-oriented wires are more prone to strain. For example tensile strain of 6% increases the SS to almost 90 mV/decade while compressive strain of 6% decreases SS to almost 65 mV/decade.

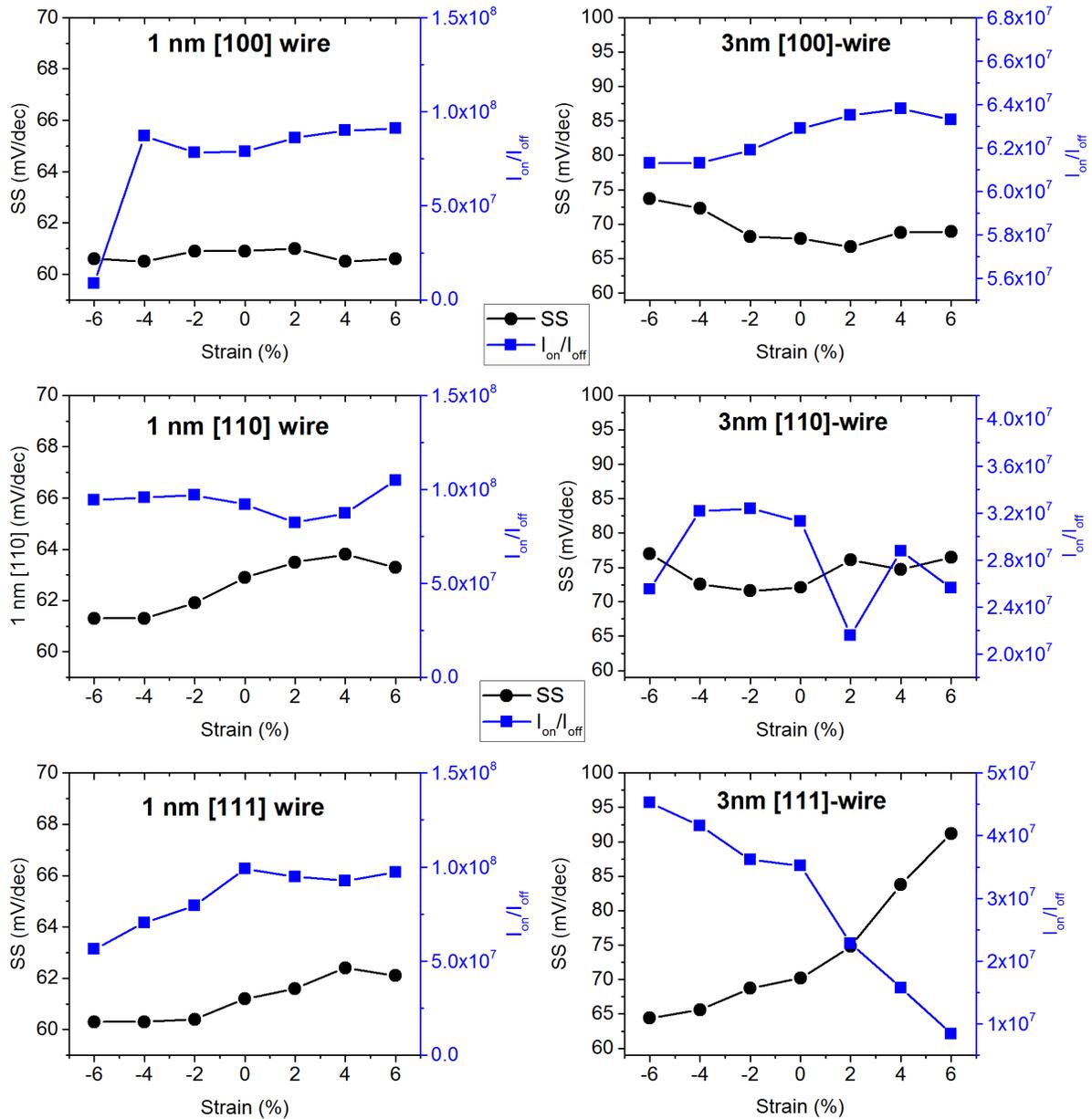

*Fig. 6.* Subthreshold swing and $I_{on}/I_{off}$ for different InAs nanowire transistors (SS is calculated at $10^{-9}$ A/μm and the current normalized by nanowire perimeter).

Fig. 6 reveals that strain has a significant effect on the electrical properties of the InAs nanowire transistors at ultra-scaled dimensions, and the effect of strain is strongly coupled to NW orientation and critical dimensions and in general increasing the SS results in decreasing of $I_{on}/I_{off}$ . Note that the simulated smaller transistors provide more $I_{on}/I_{off}$ which is due to their smaller SS due to improved electrostatic gate control.

## 5. Conclusions

We have investigated the effect of wire orientation and cross-section on the electrical properties of ultra scaled InAs nanowires with uniaxial tensile and compressive strain. Unlike bulk InAs which has an isotropic effective mass at the Γ-point, the effective masses at Γ for small diameter InAs nanowires are not isotropic due to the large quantum confinement effects. However with increasing wire diameters effective mass values of different orientations become more isotropic and return to the bulk value. Meta-GGA functionals are used for an accurate treatment of band gap energies. It is shown that the band gap and effective masses are highly dependent on nanowire diameter and orientation and this dependence has been quantified. Our results illustrate applying tensile and compressive uniaxial stress to the [111]-oriented NWs has a significant effect on the electrical properties of the InAs nanowire transistors at ultra scaled dimensions, and the effect of strain is strongly coupled to NW orientation and nanowire diameter (critical dimensions). Particularly, it was observed that the 3 nm [111]-oriented NW transistors are more prone to the effects of strain.

The results show that strain engineering can be used to improve device characteristics just as is commonly performed for silicon technologies. On the other hand, the large effect of strain on physical properties of these InAs NWs also suggests that controlling process induced strain will be critical to maintaining uniform device characteristics.

**ACKNOWLEDGMENTS**

This work was supported by the European Union project DEEPEN funded under NMR-2013-1.4-1 grant agreement number 604416. We also wish to acknowledge the SFI/HEA Irish Centre for High-End Computing (ICHEC) for the provision of computational facilities and a Science Foundation Ireland Investigator Award 13/IA/1956.

**REFERENCES**

1  Razavi, P. *et al.* Influence of channel material properties on performance of nanowire transistors. *Journal of Applied Physics* **111**, 124509, doi:10.1063/1.4729777 (2012).
2  Ansari, L., Feldman, B., Fagas, G., Colinge, J.-P. & Greer, J. C. Simulation of junctionless Si nanowire transistors with 3 nm gate length. *Applied Physics Letters* **97**, 062105-062103 (2010).

3       Ferain, I., Colinge, C. A. & Colinge, J.-P. Multigate transistors as the future of classical metal-oxide-semiconductor field-effect transistors. *Nature* **479**, 310-316 (2011).
4       Copple, A., Ralston, N. & Peng, X. Engineering direct-indirect band gap transition in wurtzite GaAs nanowires through size and uniaxial strain. *Applied Physics Letters* **100**, 193108, doi:10.1063/1.4718026 (2012).
5       Signorello, G. *et al.* Inducing a direct-to-pseudodirect bandgap transition in wurtzite GaAs nanowires with uniaxial stress. *Nature Communications* **5**, 3655, doi:10.1038/ncomms4655 http://www.nature.com/articles/ncomms4655#supplementary-information (2014).
6       Peng, X. & Copple, A. Origination of the direct-indirect band gap transition in strained wurtzite and zinc-blende GaAs nanowires: A first principles study. *Physical Review B* **87**, 115308 (2013).
7       Razavi, P. & Fagas, G. Electrical performance of III-V gate-all-around nanowire transistors. *Applied Physics Letters* **103**, 063506, doi:10.1063/1.4817997 (2013).
8       Khayer, M. A. & Roger, K. L.    (2010).
9       Tomioka, K., Yoshimura, M. & Fukui, T. A III-V nanowire channel on silicon for high-performance vertical transistors. *Nature* **488**, 189-192, doi:http://www.nature.com/nature/journal/v488/n7410/abs/nature11293.html#supplementary-information (2012).
10      Colinge, J. P. & Greer, J. C. *Nanowire Transistors: Physics of Devices and Materials in One Dimension*. (Cambridge University Press, 2016).
11      Jiang, X. *et al.* InAs/InP Radial Nanowire Heterostructures as High Electron Mobility Devices. *Nano Letters* **7**, 3214-3218, doi:10.1021/nl072024a (2007).
12      Tomioka, K., Tanaka, T., Hara, S., Hiruma, K. & Fukui, T. III–V Nanowires on Si Substrate: Selective-Area Growth and Device Applications. *IEEE Journal of Selected Topics in Quantum Electronics* **17**, 1112-1129, doi:10.1109/JSTQE.2010.2068280 (2011).
13      Niquet, Y.-M., Delerue, C. & Krzeminski, C. Effects of Strain on the Carrier Mobility in Silicon Nanowires. *Nano Letters* **12**, 3545-3550, doi:10.1021/nl3010995 (2012).
14      Zardo, I. *et al.* Pressure Tuning of the Optical Properties of GaAs Nanowires. *ACS Nano* **6**, 3284-3291, doi:10.1021/nn300228u (2012).
15      Conzatti, F., Pala, M. G., Esseni, D., Bano, E. & Selmi, L. Strain-Induced Performance Improvements in InAs Nanowire Tunnel FETs. *IEEE Transactions on Electron Devices* **59**, 2085-2092, doi:10.1109/TED.2012.2200253 (2012).
16      Cheung, H.-Y. *et al.* Modulating Electrical Properties of InAs Nanowires via Molecular Monolayers. *ACS Nano* **9**, 7545-7552, doi:10.1021/acsnano.5b02745 (2015).
17      Razavi, P. & Greer, J. C. Influence of surface stoichiometry and quantum confinement on the electronic structure of small diameter InxGa1-xAs nanowires. *Materials Chemistry and Physics* **206**, 35-39, doi:https://doi.org/10.1016/j.matchemphys.2017.12.006 (2018).
18      Ma, D., Lee, C., Au, F., Tong, S. & Lee, S. Small-diameter silicon nanowire surfaces. *Science* **299**, 1874-1877 (2003).
19      Jung, K., Mohseni, P. K. & Li, X. Ultrathin InAs nanowire growth by spontaneous Au nanoparticle spreading on indium-rich surfaces. *Nanoscale* **6**, 15293-15300 (2014).
20      Zerveas, G. *et al.* Comprehensive comparison and experimental validation of band-structure calculation methods in III–V semiconductor quantum wells. *Solid-State Electronics* **115, Part B**, 92-102, doi:http://dx.doi.org/10.1016/j.sse.2015.09.005 (2016).
21      Wu, S.-Y. *et al.* in *Electron Devices Meeting (IEDM), 2016 IEEE International.*  2.6. 1-2.6. 4 (IEEE).
22      Xie, R. *et al.* in *Electron Devices Meeting (IEDM), 2016 IEEE International.*  2.7. 1-2.7. 4 (IEEE).
23      Huynh-Bao, T. *et al.* in *IC Design & Technology (ICICDT), 2015 International Conference on.*  1-4 (IEEE).
24      Smith, T. *et al.* High bat (*Chiroptera*) diversity in the Early Eocene of India. *Naturwissenschaften* **94**, 1003-1009, doi:10.1007/s00114-007-0280-9 (2007).
25      Brandbyge, M., Mozos, J.-L., Ordejón, P., Taylor, J. & Stokbro, K. Density-functional method for nonequilibrium electron transport. *Physical Review B* **65**, 165401 (2002).
26      José, M. S. *et al.* The SIESTA method for ab initio order- N materials simulation. *Journal of Physics: Condensed Matter* **14**, 2745 (2002).
27      Tran, F. & Blaha, P. Importance of the Kinetic Energy Density for Band Gap Calculations in Solids with Density Functional Theory. *The Journal of Physical Chemistry A* **121**, 3318-3325, doi:10.1021/acs.jpca.7b02882 (2017).

28  Tao, J., Perdew, J. P., Staroverov, V. N. & Scuseria, G. E. Climbing the Density Functional Ladder: Nonempirical Meta\char21{}Generalized Gradient Approximation Designed for Molecules and Solids. *Physical Review Letters* **91**, 146401 (2003).
29  Monkhorst, H. J. & Pack, J. D. Special points for Brillouin-zone integrations. *Physical Review B* **13**, 5188-5192 (1976).
30  Abramowitz, M. *Handbook of Mathematical Functions, With Formulas, Graphs, and Mathematical Tables*. (Dover Publications, Incorporated, 1974).
31  Huang, X., Lindgren, E. & Chelikowsky, J. R. Surface passivation method for semiconductor nanostructures. *Physical Review B* **71**, 165328 (2005).
32  Skotnicki, T. & Boeuf, F. in *VLSI Technology (VLSIT), 2010 Symposium on.*  153-154.